\documentclass[5p, sort&compress]{elsarticle}

\makeatletter
\def\@author#1{\g@addto@macro\elsauthors{\normalsize%
    \def\baselinestretch{1}%
    \upshape\authorsep#1\unskip\textsuperscript{%
      \ifx\@fnmark\@empty\else\unskip\sep\@fnmark\let\sep=,\fi
      \ifx\@corref\@empty\else\unskip\sep\@corref\let\sep=,\fi
      }%
    \def\authorsep{\unskip,\space}%
    \global\let\@fnmark\@empty
    \global\let\@corref\@empty  %% Added
    \global\let\sep\@empty}%
    \@eadauthor={#1}
}

\def\ps@pprintTitle{%
 \let\@oddhead\@empty
 \let\@evenhead\@empty
 \def\@oddfoot{}%
 \let\@evenfoot\@oddfoot}
\makeatother

\usepackage{placeins}
\usepackage{lineno,hyperref}
\modulolinenumbers[5]
\usepackage{float}
\usepackage{graphicx}
\usepackage{amssymb}
\usepackage{amsmath}
\usepackage{color}
\usepackage{url}
\usepackage{threeparttable}
\usepackage{rotating}

\begin{document}

\begin{frontmatter}
\title{Revised performance parameters of the ZEPLIN-III dark matter experiment}

\author{Henrique Ara\'{u}jo\fnref{myfootnote}}
%\cortext[correspondingauthor]{Corresponding author}
\fntext[myfootnote]{{\em email}: h.araujo@imperial.ac.uk}
\author{on behalf of the ZEPLIN-III Collaboration}
%\author{A.~N.~Other}

\address{High Energy Physics Group, Department of Physics, Imperial College London, UK}





\begin{abstract}

This note presents revised detector parameters applicable to data from the First Science Run of the ZEPLIN-III dark matter experiment; these datasets were acquired in 2008 and reanalised in 2011. This run demonstrated electron recoil discrimination in liquid xenon at the level of 1 part in 10,000 below 40~keV nuclear recoil energy, at an electric field of 3.8~kV/cm; this remains the best discrimination reported for this medium to date. Building on relevant measurements published in recent years, the calibration of the scintillation and ionisation responses for both electron and nuclear recoils, which had been mapped linearly to $^{57}$Co $\gamma$-ray interactions, is converted here into optical parameters which are better suited to relate the data to the emerging liquid xenon response models. Additional information is given on the fitting of the electron and nuclear recoil populations at low energy. The aim of this note is to support the further development of these models with valuable data acquired at high field.
\end{abstract}

%\begin{keyword}
%XXX \sep XXX \sep XXX
%\MSC[July 2020]
%\end{keyword}

\end{frontmatter}

%\linenumbers

%%%%%%%%%%%%%%%%%%%%%%%%%%%%%%%%%%%%%%%%%%%%%%%%%%%%%%%%%%%%%%%%%%%%%%%%%%%
\section{Introduction}
\label{S:Introduction}

%\subsection{A subsection XXX}
%\label{S:XXX}

%\begin{figure}[ht]
%\centerline{\includegraphics[width=\linewidth]{figures/XXX.pdf}}
%\caption{\small XXX}
%\label{Fig:XXX}
%\end{figure}

The ZEPLIN-III dark matter search experiment operated at the Boulby Underground Laboratory (UK) in the period 2008--2011. This pioneering liquid xenon time projection chamber (LXe-TPC) produced competitive results for various scattering interactions~\cite{Lebedenko2009a,Lebedenko2009b,Akimov2010,Akimov2012}. In particular, it achieved the best electron recoil discrimination reported in this technology to date, operating at an electric field approaching 4~kV/cm. The First Science Run (FSR) of the experiment consisted of 67 live days of data acquired in 2008; main results were presented in Ref.~\cite{Lebedenko2009a}, hereafter `paper~1'. If the parameters set out in this note are used it would be appropriate to cite Ref.~\cite{Lebedenko2009a} to credit the experimental work as well as this note for explanation of the new parameters. The instrument was upgraded with lower background photomultiplier tubes (PMTs) for a Second Science Run (SSR), but the optical performance of the instrument was degraded (leading to poorer discrimination)~\cite{Akimov2012}. The FSR dataset was therefore the more valuable one for further studies.

At the time of those results a microscopic model for the response of liquid xenon (LXe) to low-energy particle interactions was still outstanding~\cite{Chepel2013}, and results from this and other contemporaneous experiments were not presented in a way that would make them immediately useful today, especially in what concerns energy calibration. Major progress has been made since to understand the mechanisms underlying the scintillation and ionisation responses in this technology, in terms of the respective yields and their fluctuations for electron and nuclear recoil interactions, and the discrimination between them. These enabled the development of an accurate microscopic model for the response of LXe to those interactions, incorporated into the Noble Element Simulation Technique (NEST)~\cite{NESTv2}. The behaviour of those quantities at high field is an important piece of the puzzle, yet to be fully understood~\cite{Akerib2020}.

At the core of the new microscopic model is a common “$W$-value” for the production of quanta of scintillation and ionisation for electron recoil interactions; for nuclear recoils, it is commonly considered that a Lindhard factor (${\cal L}$) multiplies the energy deposited to estimate the energy transferred to the electronic system of the colliding atoms, to which the common $W$-value then applies (see, e.g.~\cite{Akerib2018prd}. NEST can be used to predict the number of scintillation photons and ionisation electrons escaping the interaction site. In turn, the detector response to scintillation (S1) and ionisation (S2) can be predicted if the photon detection efficiencies (PDE) for the liquid and vapour phases are known: these are $g_1$ and $g_{1,gas}$, representing the number of photons detected (phd) per photon emitted in the liquid and in the gas phases, respectively; and $g_2$, representing the number of detected photons per ionisation electron escaping the primary track. These parameters are needed to relate the macroscopic S1 and S2 responses to the microscopic model.

In this note a consistent set of response parameters is derived from those published at the time, correcting the original results when more modern measurements exist. In the published analyses the two channels were calibrated relative to a reference -- the 122.1~keV and 136.5~keV $^{57}$Co $\gamma$-rays -- but relating the measured S1 and S2 responses to the microscopic model does require these additional PDE parameters. An example is the study of discrimination in LXe just published by the LUX collaboration~\cite{Akerib2020}, which includes results derived from ZEPLIN-III data in its Appendix~A.

A reanalysis of the FSR datasets in 2011 corrected a fault in the original determination of the nuclear recoil scintillation yield~\cite{Horn2011}, and improved data quality more generally through further development of the vertex reconstruction algorithm~\cite{Solovov2012}. These improved the energy resolution for $^{57}$Co $\gamma$-rays from $\sigma$=5.4\% to 3.6\%. At low energy the discrimination performance did not change appreciably. An equally important measurement was the precise determination of the single electron response, which allowed calibration of the electroluminescence response in FSR conditions~\cite{Santos2011}. In this note the new parameters are derived from the reanalysed data, although in most cases the original analysis described in paper~1 produced very similar results.

%%%%%%%%%%%%%%%%%%%%%%%%%%%%%%%%%%%%%%%%%%%%%%%%%%%%%%%%%%%%%%%%%%%%%%%%%%%%%%%%%%
\section{Geometry and electric fields}
\label{S:fields}

ZEPLIN-III was a two-electrode xenon emission detector (no gate grid), with a `planar' geometry viewed by a single PMT array immersed in the liquid phase. The FSR operating pressure was 1.65~bar ($\pm$1\%), corresponding to a liquid temperature of 174.1~K. In the active region the nominal cathode-to-anode separation was 40~mm, with the liquid level located a few mm below the anode `mirror'. The detector hardware is described in Ref.~\cite{Akimov2007}.

The gas gap between the liquid surface and the anode was 4.0$\pm$0.5~mm, and the liquid height from the cathode was 36.0$\pm$0.5~mm, with errors dominated by mechanical tolerance of the detector height. The diameter of the active region was 386~mm, with the 6.5~kg analysis fiducial volume defined by a radial cut at 150~mm and a vertical cut selecting 31.9~mm.

An updated calculation of the electric fields, using a somewhat lower permittivity $\epsilon_r$=1.85~\cite{Edwards2018,Xu2019} and considering field leakage through the wire grids~\cite{DahlThesis}, lowers the field in the gas to 7.1~kV/cm, from the original value of 7.8~kV/cm reported in paper~1. The field in the liquid is 3.8~kV/cm, close to the original 3.9~kV/cm.

The thickness of the gas layer is a key parameter in determining the gain of the S2 response. The liquid and gas heights were originally calculated from capacitance sensor measurements and the electron drift velocities in the liquid and gas phases. To understand if this can be calculated more precisely than the 0.5~mm error assigned above the drift velocity in the liquid was reassessed.

In the FSR the cathode was located at 14.1~$\mu$s drift time, with a slightly higher value of 14.3~$\mu$s reported for the single-electron dataset in~Ref.~\cite{Santos2011}. The central value produces a drift speed of 2.53$\pm$0.05~mm/$\mu$s for the nominal 36.0~mm liquid height. Taking the historical measurement of 2.83~mm/$\mu$s \cite{Miller1968}, corrected to the FSR operating temperature \cite{Benetti1993}, gives 2.65~mm/$\mu$s. Systematic effects may account for this difference, including cathode grid electrostatic deflection and the increased field near the cathode wires, both of which would increase the drift speed slightly (these can be determined with additional analysis). It is noted that somewhat lower values were reported more recently~\cite{Albert2017}, albeit for lower fields. It is concluded that the gas gap cannot be determined more precisely than the 0.5~mm uncertainty without additional data analysis.

%%%%%%%%%%%%%%%%%%%%%%%%%%%%%%%%%%%%%%%%%%%%%%%%%%%%%%%%%%%%%%%%%%%%%%%%%%%%%%%%%%
\section{Optical performance and energy calibration}
\label{S:pde}

The S1 and S2 energy scales were calibrated with $^{57}$Co $\gamma$-rays at 122.1~keV and 136.5~keV, averaging 124.5~keV into the LXe target from simulations (see paper~1). The FSR calibration constants were 11.31~keVee/nVs for the S1 channel and 0.0289~keVee/nVs for the S2 channel. In this note, as in the original publications, the unit `keVee' is referenced to the equivalent S1 or S2 signals produced by the $^{57}$Co response in each channel, for both electron and nuclear recoils. The detector flat-fielding was referred to top-centre of the liquid where most $^{57}$Co $\gamma$ rays interacted. 

In this section the photon detection efficiencies and energy thresholds are presented based on these calibration constants and additional parameters, including calculation of primary quanta with NEST~v2.0.1. In addition to the single electron response determined at the time, key new measurements include the electron emission probability from the liquid, the electroluminescence yield in the cold vapour, and the double photoelectron emission (DPE) observed in PMTs at VUV wavelengths~\cite{Faham2015}.

The average DPE probability of the ETEL~D730 PMTs was estimated to be 11\% averaged over the array, based on the comparison of the two single-electron response distributions presented in Figure~2 of Ref.~\cite{Santos2011}. Dedicated measurements of one such PMT in a recent study produced consistent results~\cite{LopezParedes2018}. In the latter reference it is noted that the novel method developed for PMT calibration in ZEPLIN-III~\cite{Neves2010} in fact measured the mean of the {\em single photon} response (47~pVs, as mentioned in paper~1) instead of the single photoelectron response measured in the traditional way, which yielded a mean pulse area of 41~pVs. This is consistent with the above DPE probability, especially if applied to the more central PMTs.

The S1 PDE is calculated as $g_1$=0.0714$\pm$0.0031~phd per photon using the scintillation yield given by NEST. This translates to an S1 light yield of 1.88~phd/keVee, which is close to the value stated in paper~1 (1.8~phe/keVee) -- noting that since the mean {\em single photon} response was used, the latter in fact represents  phd/keVee. The 4.3\% discrepancy is given as the estimated error in $g_1$.

The PDE of the gas phase can be obtained from the single electron response, which was precisely measured to have a mean of 28.25$\pm$0.25~phd~\cite{Santos2011} and width of 6.76~phd (Fano factor $F$=1.66). The number of emitted photons is calculated from the electroluminescence yield data in cold vapour from Ref.~\cite{Fonseca2004} parameterised by Eq.~(4.9) in Ref.~\cite{Chepel2013}, which results in 254~photons per emitted electron. This yields $g_{1,gas}$=0.111$\pm$0.003~phd per photon. It is noted that in other detectors the $g_1$ and $g_{1,gas}$ are very similar, but the planar geometry and all-copper construction of ZEPLIN-III motivate a larger light collection efficiency from the gas phase: this is primarily due to the focusing of the S2 light at the liquid surface and the placement of the better PMTs towards the centre of the array.

The $g_2$ value, denoting the number of photons detected per ionisation electron in the liquid, is obtained from the mean S2 response to $^{57}$Co divided by the number of ionisation electrons estimated with NEST: $g_2$=15.7$\pm$0.8~phd/e. The 5\% error is dominated by the PMT calibration.

An key parameter to obtain a bottom-up model of the S2 response is the electron emission probability -- which has not been used to calculate the main PDE parameters given above other than for consistency checks. Originally, a value approaching 80\% at the FSR field was used based on historical data~\cite{Gushchin1979}, which newer measurements confirmed to be a significant overestimation: recent measurements~\cite{Edwards2018} suggest 62\%, and the latest results decrease this to $\approx$57\%.~\cite{Xu2019}, which is the value adopted here. Although this parameter was not required in the original analysis, it was used in the optical simulation model, as mentioned below. Using the lower emission probability to calculate the number of gas electrons, and hence the number of S2 photons produced by $^{57}$Co $\gamma$-rays, we derive a $g_{1,g\!a\!s}$ which is $<$3\% away from the above value, which is a good consistency test; this discrepancy is assigned as the error in the value presented above.

A detailed optical model was developed prior to the FSR~\cite{Araujo2006}. Despite this being reasonably sophisticated, it failed to match the data especially for the S2 response. This is mostly due to the effects mentioned above, in particular the overestimation of both the field in the gas and the emission probability ($\sim$60\% combined effect). The remaining discrepancies for both S1 and S2 can be explained by the uncertainty in the reflectivity of copper at 175~nm, which remains the case.

With the new PDE parameters NEST may be used to derive accurate NR and ER energies, including the WIMP region of interest (ROI) and the detector threshold. The FSR WIMP-search region was 2--16~keVee in S1, with the lower bound corresponding approximately to the 50\% efficiency onset of the S1 channel, determined by a 3-fold coincidence level. Using NEST the WIMP ROI translates to 8.6--40.7~keV for NR and 2.4--12.3~keV for ER ($\beta$) interactions. The NR range is close to that obtained from the reanalysis of the NR scintillation yield in Ref.~\cite{Horn2011} (7--36~keV). The trigger threshold for the main datasets corresponded to 11~ionisation electrons, as stated in paper~1; this was a broad threshold onset determined by hardware. A hard analysis threshold was set at 5~nVs S2 pulse area, which translates to $\sim$4~emitted electrons ($\approx$110~phd), or 7~ionisation electrons drifting in the liquid using the above parameters (0.2~keVee in the $^{57}$Co S2 scale).

%%%%%%%%%%%%%%%%%%%%%%%%%%%%%%%%%%%%%%%%%%%%%%%%%%%%%%%%%%%%%%%%%%%%%%%%%%%%%%%%%%
\section{Discrimination}
\label{S:discrimination}

In the context of two-phase xenon detectors, ER-NR discrimination typically refers to an integrated leakage from the ER band past the mean of the NR band, with the latter defined through neutron calibration, integrated over some S1 region of interest. Modernly, this is purely a benchmark parameter that allows different datasets or experiments to be compared, but in the ZEPLIN-III FSR the ER leakage past the NR band mean was the estimator for the leading background, and hence a key parameter. In the original FSR analysis a Feldman-Cousins test statistic was applied to the ROI, and no shape information was used.

Figure~1 shows the result of the 2011 reanalysis of the FSR dataset~\cite{Jubilee2012} from which the discrimination was derived, by fitting these data -- as well as the neutron calibration -- in 1~keVee wide bins in the 2--16~keVee ROI. The S1 and S2 energy scales may be converted to phd units using the respective $^{57}$Co calibration constants and the mean PMT response of 47~pVs/phd.

\begin{figure}[ht]
\centerline{\includegraphics[width=\linewidth]{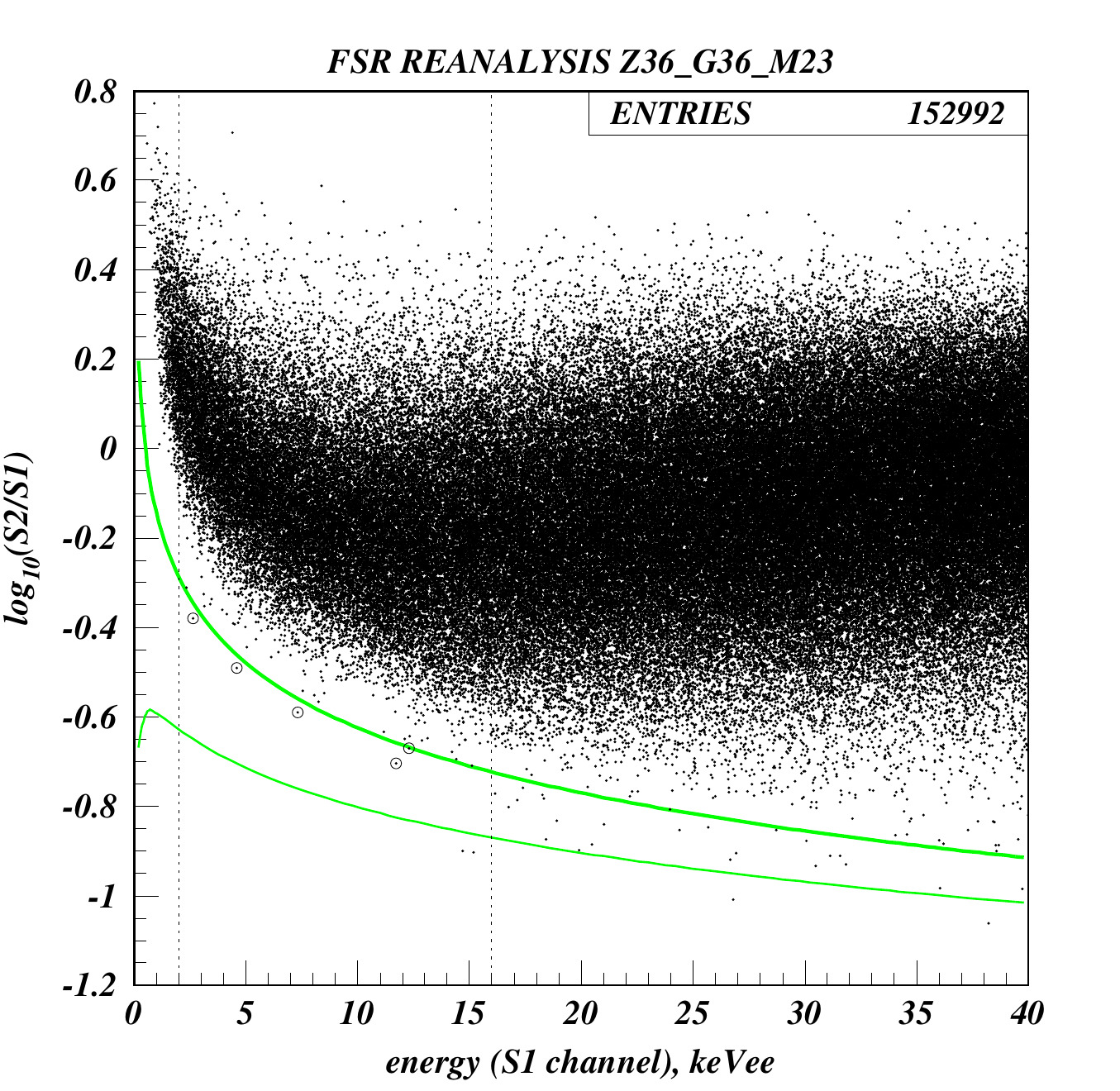}}
\caption{\small FSR dark data (2011 reanalysis) used to assess the discrimination of ER events. The WIMP-search region spanned the $\log_{10}(\mathrm{S2}/\mathrm{S1})$ range between  $\mu_{NR}$ and $\mu_{NR}-2\sigma_{NR}$ (green lines) in the range 2--16~keVee; 5~out of 38,877~events fall in this region, and a further 2~events are located below the lower boundary. At higher energies, 25~events out of 113,067 fall below the NR mean.}
\label{Fig:Scatter}
\end{figure}

The NR band was defined from Am-Be neutron calibration, with S1 bins fitted with Gaussians in $x = \log_{10}(\mathrm{S2}/\mathrm{S1})$ space. The resulting fits were parameterised in the range S1=1--40~keVee, giving an NR band mean and width equal to $\mu_{NR}=\log_{10}(0.722\,{\mathrm S1}^{-0.483})$ and $\sigma_{NR}=0.225\,{\mathrm S1}^{-0.405}$, respectively. The NR band mean and the lower 2$\sigma$ boundary are shown in the figure.

The ER band was determined from the science data shown in this same figure, which had a neutron background expectation of $\cal O$(1)~event in the ROI. A $^{137}$Cs calibration, also discussed in paper~1, was found to slightly overestimate the leakage due to PMT photocathode charging and accidental coincidences. The ER population in the `dark' data consisted mostly of low-energy Compton scatters from PMT $\gamma$-rays which dominated the background budget by a large factor.

In total, there were 38,877 events in the ROI, and the observed leakage in the reanalysed dataset was 5~events in the WIMP acceptance region (cf.~7 events in the original analysis). A further 2 events can be seen below the lower boundary; these are discussed below.

As reported in paper~1 the Skew-Gaussian distribution was found to provide a good description for shape of the ER band in $\log_{10}(\mathrm{S2}/\mathrm{S1})$ space:
\begin{displaymath}
S(x) = \frac{2A}{\omega\sqrt{2\pi}} e^{-\frac{(x-x_0)^2}{2\omega^2}}
\int_{-\infty}^{\alpha\frac{x-x_0}{w}}
\frac{1}{\sqrt{2\pi}} e^{-\frac{t^2}{2}}dt \,,
\end{displaymath}
where $x_0$ and $\omega$ denote the mean and width of the Gaussian component, $\alpha$ is the shape parameter, and $A$ is a normalisation constant. The mean and standard deviation of the distribution are $\mu=x_0+\omega\,\delta\,\sqrt{2/\pi}$ and $\sigma=\omega\sqrt{1-2\delta^2/\pi}$, with $\delta=\alpha/\sqrt{1+\alpha^2}$.

Skew-Gaussian parameters are given in Table~1 at the end of this document for fits over the range $x=[-\infty,\,\mu_{ER}+\sigma_{ER}]$. Results up to the 12--13~keVee bin are very similar if one includes the full domain; however, the quality of the fit to the critical lower tail deteriorates somewhat by forcing the upper tail to be fully described. From the fitting, the predicted leakage fraction is $1.7\times10^{-4}$, with calculated error of 70\%, which is consistent with the observed leakage of $1.3\times10^{-4}$ and 44\% Poisson error. The leakage is $1\times10^{-4}$ or lower up to $\sim$12~keVee, increasing to nearly $1\times10^{-3}$ by 15~keVee.

It is likely that a small number of events below the NR mean is not due to regular ER interactions. Aside from the small NR background from neutrons, the most challenging background topology in the experiment was due to Multiple-Scintillation Single-Ionisation (MSSI) events, which had been predicted even before the instrument had operated~\cite{Araujo2006}: the planar geometry was not favourable to reject these interactions, in which, for example, a PMT $\gamma$-ray scattered once in the LXe below the cathode and once in the active region, producing interactions with a low S2/S1 ratio. It is possible that the 2~events below the WIMP region are MSSI events. Above the ROI a total of 25 events fall below the NR mean out of a total of 113,067 events (average leakage of $2.2\times10^{-4}$); some of these are also likely MSSI events which survived the cuts developed for their removal.

For this reason, results of the fitting over the full range of $x$ are presented in Table~2. Although the fits over the restricted domain better replicated the behaviour of the leakage into the NR band, the fill fits may capture the physics of the ER band more accurately, and may be useful for that purpose.

Since discrimination is known to improve with decreasing energy in LXe (as seen in Figure~1), it is important to consider the NR spectrum generated by the neutron calibration when comparing with other experiments, in addition to differences in detector parameters such as electric field and light collection.

In a relatively large detector such as LUX, the LXe presents a thick target to MeV neutrons. At the 2.45~MeV energy of the LUX D-D generator~\cite{Akerib2016dd} the mean elastic scattering length is 19~cm, and relatively few neutrons emerge from the 50-cm diameter target without interacting elastically. In LUX, fewer than $\sim$10\% of elastic interactions are single scatters~\cite{Akerib2015bk}.

\begin{figure}[ht]
\centerline{\includegraphics[width=0.9\linewidth]{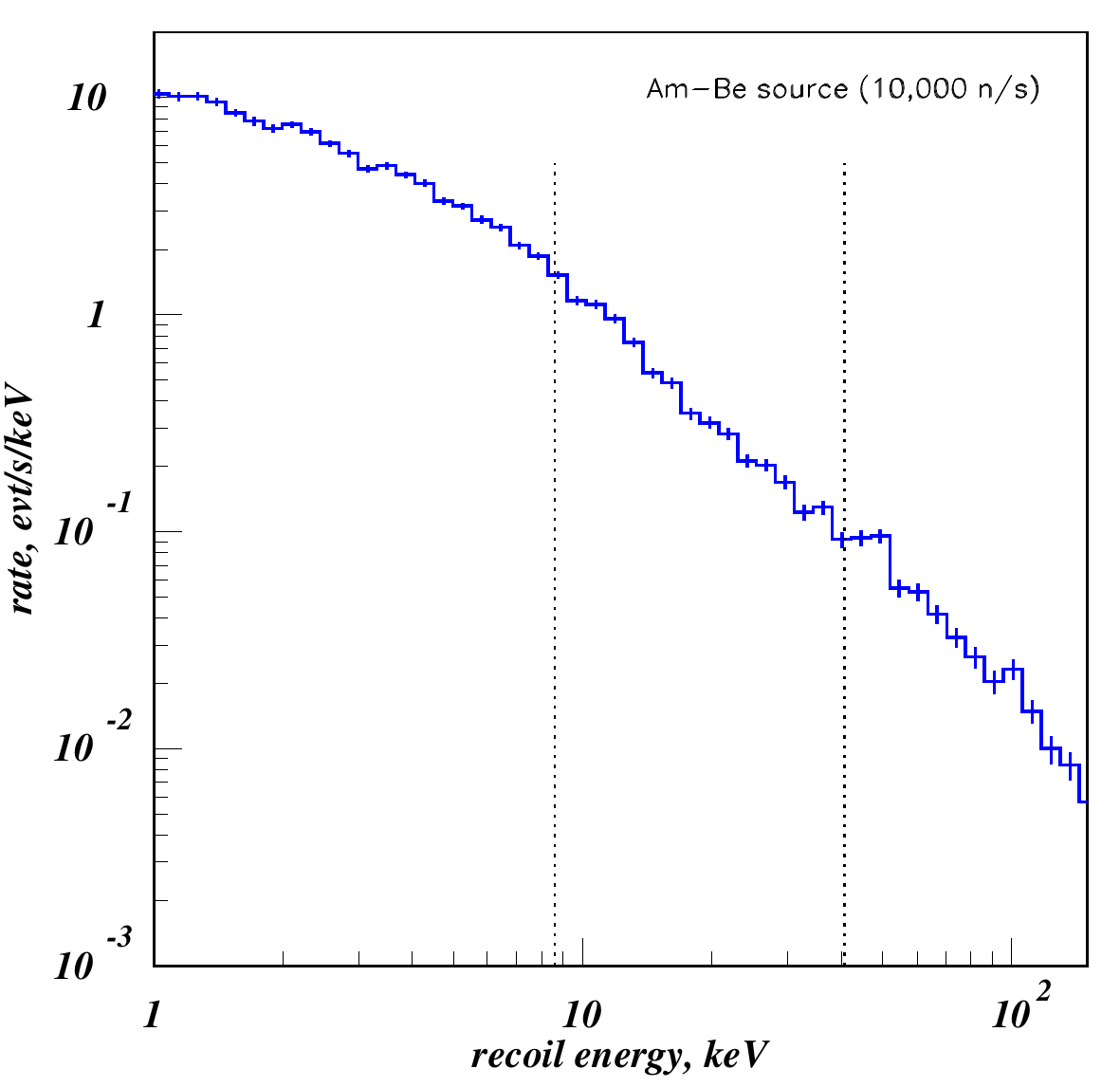}}
\caption{\small Geant4 simulation of the Am-Be calibration, showing the single-scatter recoil spectrum in the fiducial volume. This corresponds to the simulated spectrum in paper~1 but plotted here as a function of nuclear recoil energy (no energy resolution applied). The WIMP ROI is indicated by the dashed lines.}
\label{Fig:AmBecal}
\end{figure}

Contrast this with the ZEPLIN-III case. Here the neutron calibration was conducted with the Am-Be source located immediately above the detector. The mean elastic scattering length increases from $\sim$14~cm at 0.1--1~MeV neutron energy, to over 30~cm above 5~MeV. Therefore, a 4-cm thick LXe disc presents essentially as a thin target to the source. In this geometry, approximately 50\% of the elastic interactions are single scatters, with a dominant component scattering in the forward direction to produce low recoil energies --- with larger angle scatters more likely to multiple-scatter and be rejected. Figure~2 shows the simulated NR spectrum in ZEPLIN-III -- essentially the same spectrum as in paper~1. The NR spectrum in this thin detector has an enhanced rate at lower energies ($\lesssim$20~keV) due to this effect, which should be taken into account when comparing with other experiments.

\vspace{5mm}
%%%%%%%%%%%%%%%%%%%%%%%%%%%%%%%%%%%%%%%%%%%%%%%%%%%%%%%%%%%%%%%%%%%%%%%%%%%%%%%%%%%%%%%%%%%%%%%%%%%%
\section*{Acknowledgements}

Credit for the data described in this note is due to the ZEPLIN-III collaboration, which included Imperial College London (UK), the STFC Rutherford Appleton Laboratory (UK), Edinburgh University (UK), ITEP-Moscow (RU) and LIP-Coimbra (PT), working at the Boulby Underground Laboratory (UK). I am particularly grateful to Matthew Szydagis (SUNY University at Albany) for many discussions leading to his incorporating of this information into NEST, and for his heroic efforts to develop an accurate LXe response model. Thanks are also due to Vetri Velan (U.C. Berkeley) and to Ekaterina Kozlova (ITEP) for applying the new NEST models to the ZEPLIN-III discrimination.

\newpage

\begin{table*}
\caption{ER and NR band data from the ZEPLIN-III First Science Run 2011 reanalysis, with the ER band fitted only for $x = \log_{10}(\mathrm{S2}/\mathrm{S1}) \le \mu_{ER}+\sigma_{ER}$. For each 1-keVee wide bin the nuclear recoil energy (NEST~v2.0.1) and S1 photons detected are presented at the bin centre. The next set includes the number of ER counts in the bin ($N$), the observed leakage ($N_O$), the estimated leakage ($N_E$) and associated error ($\sigma_E$). Combined values for the whole ROI are presented in the last row. The ER band fit data include the four Skew-Gaussian parameters, as well as the mean and standard deviation ($\mu_{ER}$ and $\sigma_{ER}$). Corresponding parameters for the parameterisation of the Gaussian fits to the NR band are presented ($\mu_{NR}$ and $\sigma_{NR}$).\label{tab:ERfits1}}
\begin{center}
{\small
\begin{threeparttable}
\begin{tabular}{c c c | c c r r | r r r r r r | r r}
\hline\hline
\multicolumn{3}{c}{S1 bins} & \multicolumn{4}{c}{ER counts \& leakage} & \multicolumn{6}{c}{ER band: fit for $-\infty< x \le \mu_{ER} + \sigma_{ER}$} & \multicolumn{2}{c}{NR band}\\
\hline
keVee & keVnr & phd &$N$ &$N_O$ &$N_E$ &$\sigma_E$ & $A$ &$x_0$ &$\omega$ &$\alpha$ &$\mu_{ER}$ &$\sigma_{ER}$ &$\mu_{NR}$ & $\sigma_{NR}$  \\
\hline
2--3   &10.1 &4.7  &2,062 &1 &0.12 &0.42 &83.1  &$+$0.021 &0.159 &1.14 &$+$0.116 &0.127 & $-$0.333 & 0.155\\
3--4   &12.9 &6.6  &2,371 &0 &0.09 &0.31 &107.9 &$-$0.113 &0.244 &2.59 &$+$0.069 &0.163 & $-$0.404 & 0.136\\
4--5   &15.5 &8.5  &2,410 &1 &0.21 &0.32 &101.9 &$-$0.160 &0.210 &1.99 &$-$0.010 &0.147 & $-$0.457 & 0.123\\
5--6   &17.9 &10.4 &2,472 &0 &0.29 &0.54 &101.3 &$-$0.196 &0.187 &1.64 &$-$0.069 &0.137 & $-$0.499 & 0.113\\
6--7   &20.3 &12.2 &2,511 &0 &0.31 &0.42 &104.8 &$-$0.242 &0.208 &1.94 &$-$0.094 &0.147 & $-$0.534 & 0.106\\
7--8   &22.6 &14.1 &2,634 &1 &0.24 &0.34 &109.9 &$-$0.278 &0.230 &2.29 &$-$0.110 &0.157 & $-$0.564 & 0.100\\
8--9   &24.9 &16.0 &2,703 &0 &0.31 &0.55 &111.0 &$-$0.287 &0.203 &1.82 &$-$0.145 &0.145 & $-$0.590 & 0.095\\
9--10  &27.1 &17.9 &2,668 &0 &0.28 &0.37 &113.9 &$-$0.326 &0.244 &2.39 &$-$0.146 &0.165 & $-$0.614 & 0.091\\
10--11 &29.3 &19.8 &2,911 &0 &0.13 &0.26 &137.0 &$-$0.360 &0.316 &3.52 &$-$0.117 &0.203 & $-$0.635 & 0.087\\
11--12 &31.3 &21.6 &3,046 &1 &0.20 &0.32 &137.8 &$-$0.384 &0.310 &3.40 &$-$0.147 &0.199 & $-$0.654 & 0.084\\
12--13 &33.5 &23.5 &3,176 &1 &0.53 &0.55 &142.7 &$-$0.380 &0.303 &2.82 &$-$0.152 &0.200 & $-$0.671 & 0.081\\
13--14 &35.6 &25.4 &3,075 &0 &1.10 &2.25 &151.3 &$-$0.394 &0.364 &3.09 &$-$0.117 &0.237 & $-$0.687 & 0.079\\
14--15 &37.6 &27.3 &3,298 &0 &1.56 &2.36 &162.8 &$-$0.370 &0.292 &2.09 &$-$0.160 &0.202 & $-$0.702 & 0.076\\
15--16 &39.7 &29.2 &3,540 &0 &1.35 &3.17 &167.0 &$-$0.393 &0.351 &2.68 &$-$0.131 &0.233 & $-$0.716 & 0.074\\
\hline
2--16 &8.6--40.7 &3.8--30.1 &38,877 &5 &6.73 & 4.75 & -- & -- & -- & -- & -- & -- & -- & -- \\
\hline\hline
\end{tabular}
%\begin{flushleft} 
%{\footnotesize $^\dagger$ XXX} 
%\end{flushleft}
\end{threeparttable}
}
\end{center}
\end{table*}

\begin{table*}
\caption{ER and NR band data as in Table~1, but with the ER band fitted over all values of $x=\log_{10}(\mathrm{S2}/\mathrm{S1})$.\label{tab:ERfits2}}
\begin{center}
{\small
\begin{threeparttable}
\begin{tabular}{c c c | c c r r | r r r r r r | r r}
\hline\hline
\multicolumn{3}{c}{S1 bins} & \multicolumn{4}{c}{ER counts \& leakage} & \multicolumn{6}{c}{ER band: $-\infty < x < +\infty$} & \multicolumn{2}{c}{NR band}\\
\hline
keVee & keVnr & phd &$N$ &$N_O$ &$N_E$ &$\sigma_E$ & $A$ &$x_0$ &$\omega$ &$\alpha$ &$\mu_{ER}$ &$\sigma_{ER}$ &$\mu_{NR}$ & $\sigma_{NR}$  \\
\hline
2--3   &10.1 &4.7  &2,062 &1 &0.10 &0.10 &85.9  &$+$0.007 &0.178 &1.47 &$+$0.125 &0.134 & $-$0.333 & 0.155\\
3--4   &12.9 &6.6  &2,371 &0 &0.22 &0.31 &98.9  &$-$0.084 &0.187 &1.58 &$+$0.042 &0.138 & $-$0.404 & 0.136\\
4--5   &15.5 &8.5  &2,410 &1 &0.23 &0.15 &100.3 &$-$0.156 &0.201 &1.85 &$-$0.015 &0.143 & $-$0.457 & 0.123\\
5--6   &17.9 &10.4 &2,472 &0 &0.22 &0.18 &102.9 &$-$0.207 &0.203 &1.96 &$-$0.063 &0.143 & $-$0.499 & 0.113\\
6--7   &20.3 &12.2 &2,511 &0 &0.30 &0.19 &104.6 &$-$0.243 &0.209 &1.97 &$-$0.094 &0.147 & $-$0.534 & 0.106\\
7--8   &22.6 &14.1 &2,634 &1 &0.33 &0.28 &109.8 &$-$0.268 &0.218 &2.01 &$-$0.113 &0.152 & $-$0.564 & 0.100\\
8--9   &24.9 &16.0 &2,703 &0 &0.25 &0.17 &112.5 &$-$0.294 &0.215 &2.04 &$-$0.140 &0.150 & $-$0.590 & 0.095\\
9--10  &27.1 &17.9 &2,668 &0 &0.32 &0.20 &111.0 &$-$0.321 &0.229 &2.16 &$-$0.154 &0.158 & $-$0.614 & 0.091\\
10--11 &29.3 &19.8 &2,911 &0 &0.37 &0.30 &121.3 &$-$0.332 &0.236 &2.14 &$-$0.162 &0.163 & $-$0.635 & 0.087\\
11--12 &31.3 &21.6 &3,046 &1 &0.46 &0.36 &126.9 &$-$0.361 &0.249 &2.31 &$-$0.179 &0.170 & $-$0.654 & 0.084\\
12--13 &33.5 &23.5 &3,176 &1 &0.94 &0.59 &132.4 &$-$0.355 &0.245 &1.94 &$-$0.181 &0.173 & $-$0.671 & 0.081\\
13--14 &35.6 &25.4 &3,075 &0 &3.15 &3.75 &128.2 &$-$0.318 &0.220 &1.16 &$-$0.186 &0.175 & $-$0.687 & 0.079\\
14--15 &37.6 &27.3 &3,298 &0 &2.41 &1.53 &137.3 &$-$0.338 &0.244 &1.44 &$-$0.178 &0.185 & $-$0.702 & 0.076\\
15--16 &39.7 &29.2 &3,540 &0 &2.99 &3.01 &147.6 &$-$0.332 &0.238 &1.26 &$-$0.183 &0.185 & $-$0.716 & 0.074\\
\hline
2--16 &8.6--40.7 &3.8--30.1 &38,877 &5 &12.3 & 5.14 & -- & -- & -- & -- & -- & -- & -- & -- \\
\hline\hline
\end{tabular}
%\begin{flushleft} 
%{\footnotesize $^\dagger$ XXX} 
%\end{flushleft}
\end{threeparttable}
}
\end{center}
\end{table*}

%\FloatBarrier
%\section*{References}

\bibliography{mybibfile}

\end{document}